\documentclass[aps,prb,reprint,superscriptaddress]{revtex4-2}

\usepackage[utf8]{inputenc}
\usepackage[T1]{fontenc}
\usepackage{subcaption}
\usepackage{amssymb,amsmath, braket, ulem}
\usepackage{graphicx}
\usepackage{xcolor,tikz}
\usepackage[pdfusetitle]{hyperref}
\usepackage{booktabs} % in preamble
\usepackage[justification=justified,singlelinecheck=false]{caption}
\usepackage{placeins}

\usetikzlibrary{calc, positioning, arrows.meta, decorations.pathreplacing}
\definecolor{solyellow}{HTML}{b58900}
\definecolor{solorange}{HTML}{cb4b16}
\definecolor{solred}{HTML}{dc322f}
\definecolor{solmagenta}{HTML}{d33682}
\definecolor{solviolet}{HTML}{6c71c4}
\definecolor{solblue}{HTML}{268bd2}
\definecolor{solcyan}{HTML}{2aa198}
\definecolor{solgreen}{HTML}{859900}

\definecolor{vir1}{HTML}{440154} % dark purple
\definecolor{vir2}{HTML}{472c7a} % purple
\definecolor{vir3}{HTML}{3b518b} % blue
\definecolor{vir4}{HTML}{2c718e} % blue
\definecolor{vir5}{HTML}{21908d} % blue-green
\definecolor{vir6}{HTML}{27ad81} % green
\definecolor{vir7}{HTML}{5cc863} % green
\definecolor{vir8}{HTML}{aadc32} % lime green
\definecolor{vir9}{HTML}{fde725} % ye1llow

\begin{document}

\title{Ground-state estimation of the Heisenberg model on frustrated lattices with Sample-based Krylov Quantum Diagonalization}

\author{Calvin Brooks}
\affiliation{Rensselaer Polytechnic Institute, Troy, NY 12810, USA}

\author{Henry Zou}
\affiliation{IBM Quantum, IBM Research, Cambridge, MA 02142, USA}

\author{Trevor David Rhone}
\affiliation{Rensselaer Polytechnic Institute, Troy, NY 12810, USA}

\date{May 28, 2026}

\begin{abstract}

Quantum spin simulations of frustrated lattices remain challenging for both classical and quantum algorithms, particularly in parameter regimes relevant to quantum spin liquid (QSL) phases. In this work, we apply Sample-based Krylov Quantum Diagonalization (SKQD) to estimate the ground state of the antiferromagnetic XXZ Heisenberg model on the $J_1$--$J_2$ square lattice, the Kagome lattice, and a 1D chain, studying system sizes from 12 to 72 spins. In our application of SKQD, we identify a ZZ deformation of $\Delta=2$ as a sufficiently sparse Hamiltonian and introduce two modifications to the SKQD framework tailored to spin models: a canonical bitstring compression scheme that preserves the effectiveness of configuration recovery under spin-flip degeneracy, and the use of multiple Krylov subspaces to improve ground state coverage without any increase in quantum resources. For the 1D chain and Kagome lattice, SKQD achieves sub-percent ground-state energy errors at system sizes up to 24 spins, including a relative error of $0.002\%$ on the 12-site Kagome lattice, surpassing the best prior VQE result of $0.01\%$ on the same system while requiring no variational optimization. SKQD further extends to system sizes well beyond the reach of prior quantum algorithm studies, reaching 72 spins across all three geometries. Beyond 24 spins, accuracy degrades to relative errors of $19\%$--$36\%$ at 72 sites, but the gradual scaling of error with system size suggests these limits are set by available shot budgets and circuit depth rather than fundamental algorithmic constraints. Although classical tensor network methods remain state-of-the-art for these models, this work establishes a new benchmark for quantum simulation of the frustrated Heisenberg model and demonstrates SKQD as a scalable, hardware-compatible approach for studying strongly correlated spin systems.

\end{abstract}

\maketitle

\section{Introduction}\label{sec:intro}

Quantum spin models are a cornerstone of condensed matter physics, providing a tractable framework for studying strongly correlated systems and exotic phases of matter that resist description by conventional single-particle theories. Among these, the Heisenberg model has emerged as the canonical setting for investigating quantum spin liquids (QSLs) \cite{QSL_overview}, which are highly entangled phases in which magnetic order is suppressed even at zero temperature by quantum fluctuations. QSLs are of broad scientific interest due to their proposed connections to high-temperature superconductivity, topological quantum computing, and novel data storage paradigms \cite{topological_QC, datastorage_superconductivity}.

A necessary ingredient for QSL formation is geometric or exchange frustration, which prevents the system from simultaneously minimizing all pairwise interactions and suppresses conventional magnetic ordering. Frustration arises naturally from the lattice geometry in the Kagome lattice \cite{kagome_QSL}, or through competing nearest- and next-nearest-neighbor exchange interactions, as in the $J_1$--$J_2$ square lattice \cite{square_QSL_evidence}. The $J_1$--$J_2$ square lattice is of particular interest near $J_2/J_1 \approx 0.5$, where frustration is maximal and a QSL phase has been proposed \cite{square_QSL_evidence, QSL_found}, though its existence and character remain contested \cite{QSL_absence}. The Kagome lattice is similarly debated, with strong numerical evidence for a gapless QSL ground state \cite{kagome_QSL}. In this work, we study the XXZ Heisenberg model(Eq.~\ref{eqn:XXZ_heisen_ham}) on both of these geometries, along with a 1D chain as a 
validation case.

\begin{figure*}[t]
    \centering
    \includegraphics[width=\textwidth]{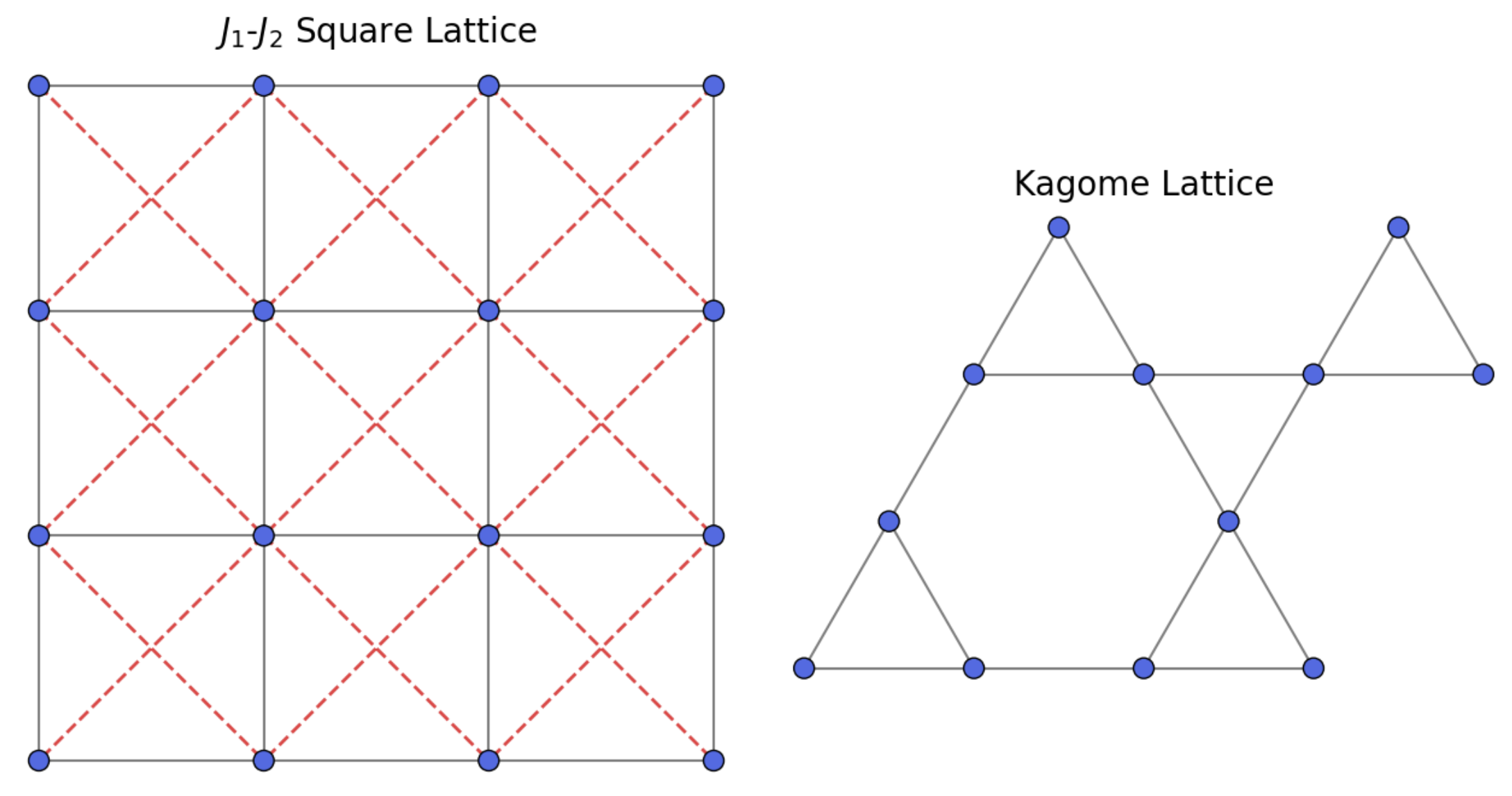}
    \caption{Examples of lattices studied in this work. (Left) $4\times4$ square lattice.  Nearest-neighbor ($J_1$) interactions are shown as solid black lines and next-nearest-neighbor ($J_2$) interactions as dashed red lines. (Right) $2\times2\times3$ Kagome lattice. The unit cell consists of three sites arranged in a triangle, tiled into 
    a $2\times2$ supercell.}
    \label{fig:lattice_types}
\end{figure*}

The ground state of the Heisenberg model remains the object of extensive study. Several classical methods have been developed for estimating the ground state of 2-dimensional lattices. Exact diagonalization quickly becomes intractable as system size increases due to the exponential growth of the Hilbert space, leading to limits of $\sim 40$ spins \cite{exact_diagonalization}. Beyond exact methods, several tensor network approaches have been applied to the Heisenberg model, including Density Matrix Renormalization Group (DMRG) \cite{Triangle_QSL_DMRG} and Projected Entangled Pair States (PEPS) \cite{Square_PEPS}. Although these techniques have achieved considerable success, scaling issues remain due to heavy entanglement. In particular, DMRG struggles with non-ribbon-like lattices \cite{2D_DMRG}, and PEPS requires extremely taxing tensor contraction on lattices with long-range entanglement, usually caused by periodic boundary conditions \cite{PEPS_PBC}. Additionally, these methods give inconsistent reports regarding the existence and stability of the QSL phase in certain parameter regimes~\cite{QSL_absence, QSL_found}.

Quantum computers offer a qualitatively different route, as they can in principle represent and manipulate the full quantum state without truncations or ansatz-dependent approximations, making them a compelling platform for probing strongly correlated phases such as QSLs. Previous quantum algorithm studies have applied the Variational Quantum Eigensolver (VQE) to frustrated Heisenberg models. Weaving et al.\ achieved a ground-state energy error of $0.01\%$ on a 12-site Kagome system \cite{kagome_12spin}, demonstrating that quantum methods can reach high accuracy on small systems. However, VQE does not scale to larger systems: the required circuit depth grows rapidly with system size, and the barren plateau problem causes gradient-based optimization to become exponentially harder as the number of qubits increases \cite{VQE_review}. Shiels (2024) encountered this limitation directly, finding that VQE energy estimates on the $J_1$--$J_2$ square lattice became unreliable beyond 16 spins, shifting the focus of that work to alternative observables \cite{Shiels_16spin_square}.

Sample-based Krylov Quantum Diagonalization (SKQD) \cite{SKQD} offers an alternative that is better suited to near-term hardware. Rather than optimizing a variational circuit, SKQD constructs a Krylov subspace from time-evolved measurement samples and diagonalizes the Hamiltonian classically within that subspace, using a configuration recovery step to iteratively refine noisy bitstring data. This separation of quantum sampling from classical diagonalization sidesteps the barren plateau problem and is more tolerant of hardware noise than deep variational circuits.

In this work, we apply SKQD to estimate the ground state of the XXZ Heisenberg model (Eq.~\ref{eqn:XXZ_heisen_ham}) on the $J_1$--$J_2$ square and Kagome lattices, along with a 1D chain, studying system sizes from 12 to 72 spins on \texttt{ibm\_pittsburgh} (Heron~r3). We introduce two modifications to the SKQD framework that are essential for reliable performance on spin models: a canonical bitstring compression scheme that preserves the effectiveness of configuration recovery under spin-flip degeneracy, and multi-initial state Krylov circuit to improve subspace coverage while not increasing quantum overhead. Our results surpass prior quantum algorithm studies in both accuracy and system size. For the 12-site Kagome lattice, SKQD achieves a ground-state energy error of $0.0020\%$, outperforming the best prior VQE result of $0.01\%$ on the same system, and sub-percent accuracy is maintained up to 24 spins. While classical tensor network methods remain state-of-the-art for these models, this work establishes a new benchmark for quantum simulation of the frustrated Heisenberg model.

\begin{equation}
\begin{aligned}
\hat{H}_{J_1J_2} &=
J_1 \sum_{\langle i,j\rangle}\left(X_iX_j+Y_iY_j+\Delta Z_iZ_j\right) \\
&\quad + J_2 \sum_{\langle\!\langle i,j\rangle\!\rangle}
\left(X_iX_j+Y_iY_j+\Delta Z_iZ_j\right)
\end{aligned}
\label{eqn:XXZ_heisen_ham}
\end{equation}

During the preparation of this work, a related study by Firt et al. \cite{SKQD_on_heisenberg} appeared, which investigates the performance of SKQD in Heisenberg models, focusing mainly on 1D chains and unfrustrated 2D lattices in the presence of an external field. However, extending SKQD to frustrated models without an external field introduces challenges related to ground state sparsity and degeneracy within the ground state. In this work, we target this regime by introducing several modifications to the SKQD framework.

\section{Methods}\label{sec:methods}

Sample-based Krylov Quantum Diagonalization (SKQD) \cite{SKQD} is an iterative algorithm that refines noisy measurement outcomes from a set of Krylov circuits that are run on a quantum computer. It is based on Sample-based Quantum Diagonalization (SQD) \cite{Robledo_Moreno_2025}, which first introduced the Configuration Recovery process. In each iteration, the ground state is estimated, this estimate is used to recover noisy bitstrings from the input data, and the expanded dataset is fed back to refine the ground state in the next iteration. This loop continues until convergence. SQD was originally proposed for electronic structure problems; here, we introduce several modifications to improve performance for quantum spin models.

\subsection{Projected Hamiltonian and Occupancy Estimation}

The ground state is estimated by projecting the Hamiltonian into the subspace spanned by the collected bitstrings. The projected Hamiltonian is diagonalized classically, yielding an approximation to the true ground state.

After diagonalization, the artificially added complementary bitstrings (discussed in section C) are removed to reduce computational overhead, as they will be reintroduced before the next projection step. The average spin occupancy is then computed by squaring the components of the ground state eigenvector and taking the expectation value over the bitstring basis. This produces a $1 \times N$ vector, where each entry represents the probability that the corresponding spin is in the $\ket{1}$ state.

If complementary bitstrings were retained during this step, the average occupancy would be forced to $0.5$ for every spin due to symmetry. Removing complements allows the occupancy vector to retain physically meaningful information.

\subsection{Configuration Recovery}

The average occupancy information is used to refine the input bitstrings. Due to hardware noise, some measured bitstrings may be unphysical, meaning they violate spin conservation. For example, the ground state of a four-spin XXZ chain lies in the $S_z = 0$ sector (two spins up and two spins down). A measured bitstring such as 1110 violates this constraint and is therefore noise-induced. Rather than discarding these states, we probabilistically flip each bit according to the estimated average occupancy. Since this occupancy is derived from the previous ground state estimate, unphysical bitstrings that are close to the true ground state can be recovered and incorporated into the next iteration. As a novel modification to the algorithm, the top 30\% of recovered bitstrings at each iteration are retained and explicitly inserted into the next cycle to encourage convergence. A detailed description of this procedure can be found in the supplementary material of Ref.~\cite{Robledo_Moreno_2025}. 

In this work, Hamiltonian projection, diagonalization, and configuration recovery are implemented using the Qiskit SQD add-on \cite{qiskit-addon-sqd}.

\subsection{Spin Filtering and Degeneracy Expansion}

The raw quantum circuit output consists of a list of measured bitstrings. We first filter these bitstrings to include only those with total spin $S_z = 0$, corresponding to a Hamming weight equal to $N/2$ for a system of $N$ spins. For antiferromagnetic models, the ground state lies in the spin-0 sector, and therefore only bitstrings satisfying this constraint contribute to the true ground state.

Due to the symmetry of the Heisenberg Hamiltonian, the total energy depends only on spin interactions and not on the global spin inversion. Consequently, bitwise complements (e.g., 1001 and 0110) have identical energies and contribute equally to the ground state. We exploit this degeneracy by explicitly adding each complement to the dataset prior to projection and diagonalization. This increases the effective coverage of the relevant subspace and reduces the burden on the configuration recovery step.

An important caveat is the configuration recovery (CR) process and its weakness to this degeneracy. CR removes noise by constructing an estimate of the ground state as an average occupancy array. This is formed by taking an average of all bitstrings in the previous eigensolve, each weighted by their coefficient in the ground state eigenvector. Since degenerate bitstrings contribute equally to the ground state, this forces the average occupancy array to become $[0.5, 0.5, 0.5, 0.5, \dots]$, invalidating CR. To avoid this issue, we compress the bitstring dataset before we calculate this average occupancy by 
pruning one element of every complement pair in the set.
%removing one out of every complement pair.
This creates a ``canonical'' dataset, which we can use to create a meaningful average occupancy array to use for CR. We arbitrarily choose to remove the bitstring with the leading 1 when forming this dataset.

\subsection{Quantum Circuit Construction and Krylov Expansion}

The difference between SQD and SKQD lies in the quantum circuits used to construct our subspace for projecting the Hamiltonian. While the original implementation of SQD used the LUCJ ansatz for electronic structure, we instead build these measurement data as a Krylov subspace using time evolution operators.

Krylov methods are well-established classical techniques for eigenvalue estimation. Jiang et al.~\cite{Yoshioka2025} demonstrated that Krylov-based constructions can be realized in quantum settings via time evolution. In our approach, the measurement data generated from this Krylov expansion is fed into the SKQD algorithm described above. The Krylov space is defined in Eqn.~\ref{eqn:krylov}, where $U = e^{-iHk\Delta t}$ and the timestep is set to $\Delta t = \pi / \sum_j |c_j|$, with $\sum_j |c_j|$ serving as an upper bound on the spectral norm of $H = \sum_j c_j P_j$.This ensures the phase accumulated over one timestep satisfies $\|H\|\Delta t \lesssim \pi$. If $\Delta t$ were too small, successive Krylov vectors would be nearly identical and the subspace would be poorly conditioned; if too large, the evolution would over-rotate and lose its connection to the low-energy structure of $H$. This choice therefore balances non-trivial evolution against spectral resolution of the ground state. 

Running circuits with different values of $k$ is equivalent to repeatedly applying $U$. By aggregating measurement data from all such circuits, we effectively build a Krylov subspace.
\begin{equation}
    K^r = \{\ket{v}, U\ket{v}, U^2\ket{v}, ..., U^{r-1}\ket{v}\}
    \label{eqn:krylov}
\end{equation}  

\begin{figure}
    \centering
    \includegraphics[width=0.65\linewidth]{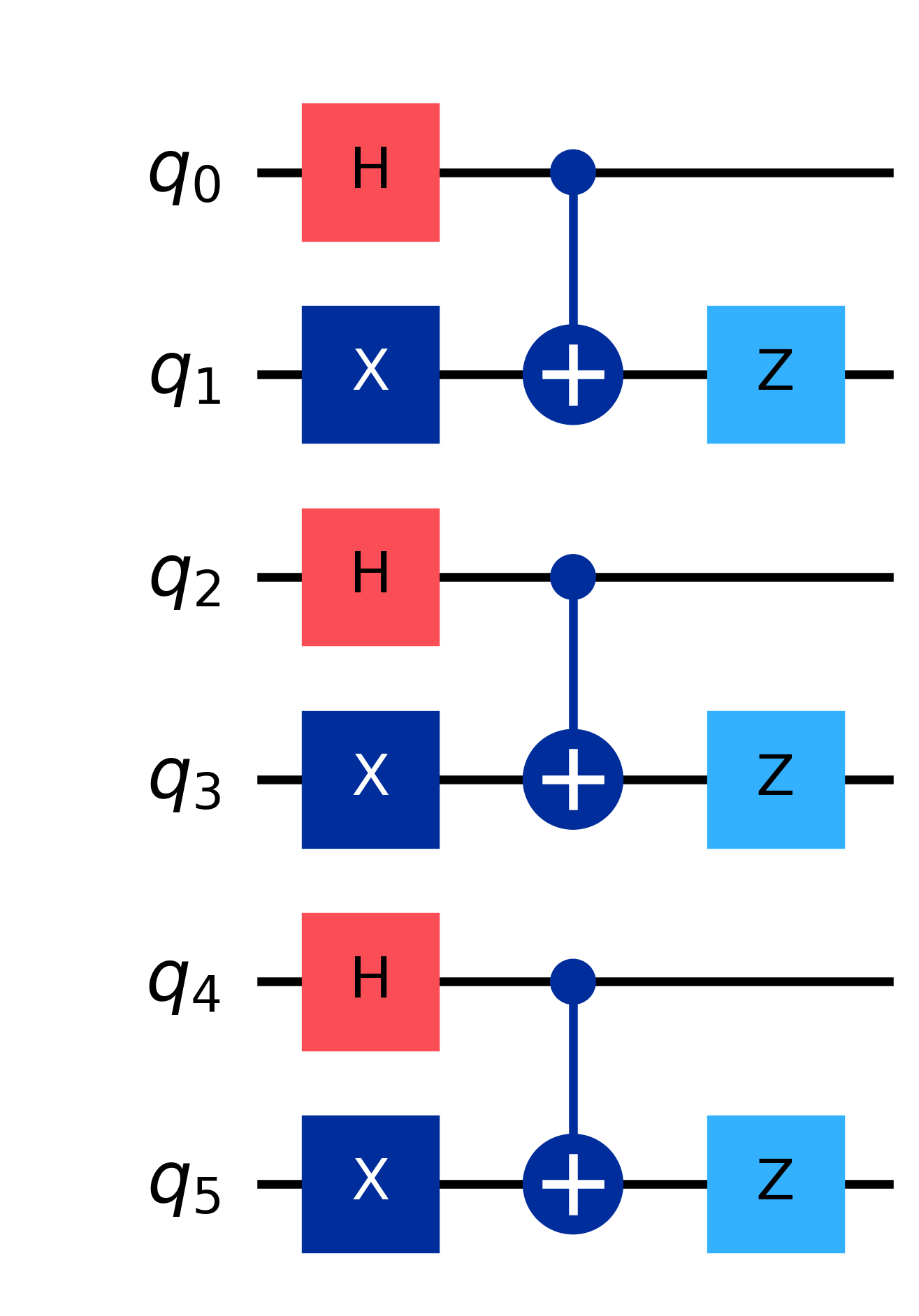}
    \caption{Example of Dimer/singlet initial state preparation on 6 qubits.}
    \label{fig:initial_states}
\end{figure}

The initial state/vector $v$ should lie in the same spin sector as the ground state. For the antiferromagnetic XXZ model, we use two initial states and combine the output to create a larger Krylov space. The first state used is a Dimer state, also known as a singlet state. This involves initializing pairs of qubits into the entangled state $\frac{1}{\sqrt{2}}(\ket{01} - \ket{10})$ (see Fig.~\ref{fig:initial_states}). This lies in the spin-0 subspace, making it ideal for our problem. The second state is a warm-start, obtained by applying a variational mean-field (product-state) approximation \cite{warm_start}. We start our approximation in the Néel state, then perform sweeps across the lattice and optimize single-qubit Bloch angles with coordinate descent to find the lowest energy. Sweeps are continued until no energy improvements are found. We then implement this state using an $R_z$ and $R_y$ gate applied to each qubit. Although this initial state contains no entanglement information, it provides a lower-energy and more physically informed starting point than simple product states such as Néel. Using two initial states encourages a diverse Krylov subspace which becomes necessary when dealing with larger systems beyond $\sim40 $ spins. 

The actual form of this Krylov subspace in our work is the set of bitstrings that we measure from each of these circuits. Therefore, the dimension of the subspace is not equal to $r$, but is instead equal to the total number of bitstrings we collect with duplicates removed. We then project the Hamiltonian into the subspace defined by these bitstrings (Eqn.~\ref{eqn:projection}), find the ground state of the projected Hamiltonian $\tilde{H}$ classically, and use this as an estimate of the true ground state. 

\begin{equation}
    \tilde{H}_{ij} = \bra{\psi_i}H\ket{\psi_j} \quad, \quad \psi_i = 01110110001\dots
    \label{eqn:projection}
\end{equation}

Due to hardware constraints, the number of Krylov steps is limited by circuit depth and noise. To enable a consistent comparison across lattice geometries, we fix the maximum Krylov order to k=5 for the two-dimensional lattice instances. This choice was guided by the transpiled two-qubit-gate depth of the most demanding geometry considered, the $J_1-J_2$ square lattice. For this lattice, k=5 gives average two-qubit depths ranging from approximately 231 for 12 spins to 676 for 72 spins, placing the largest circuits near our intended hardware-feasibility scale of $\sim$600 two-qubit layers. This scale was determined based on guidance from IBM collaborators. The same Krylov order is therefore used for the kagome lattice, whose corresponding depths are lower, ranging from approximately 104 to 281 over the same system-size range. The one-dimensional chain is substantially shallower and is studied up to k=20, with an average two-qubit depth of approximately 63 across the system sizes considered.

\subsection{Classical Baselines}

Two methods were used to obtain reference energies for benchmarking our SKQD results. The first is exact diagonalization (ED), in which the Hamiltonian is
projected into the full $S_z = 0$ Hilbert space and the ground-state energy is computed directly via sparse eigensolvers, as implemented in the Qiskit SQD
add-on~\cite{qiskit-addon-sqd}. Because the dimension of the spin-0 subspace grows as $\binom{N}{N/2}$, exact diagonalization is limited to systems of at most 24 spins on the hardware available to us.

For larger systems, we use the Density Matrix Renormalization Group (DMRG) ~\cite{white1992dmrg, white1993dmrg}. DMRG represents the quantum state as a Matrix Product State (MPS), a one-dimensional tensor network in which each site is described by a tensor of bounded dimension $\chi$, called the bond dimension. The algorithm variationally optimizes the MPS tensors sweep by sweep to minimize the energy expectation value, keeping only the $\chi$ most significant singular values at each bond. For 1D systems, the entanglement in low-energy states obeys an area law, meaning the entanglement entropy of any bipartition is bounded by a constant, so a modest bond dimension suffices for essentially exact results. In this work, we apply DMRG to 1D XXZ Heisenberg chains using a schedule of increasing bond dimensions with a truncation cutoff of $10^{-10}$, obtaining ground-state energies that serve as accurate benchmarks for our chain results.

Applying DMRG to two-dimensional lattices such as the Kagome lattice requires additional care. Because DMRG operates on a 1D tensor network, a 2D geometry must first be mapped onto a linear chain ordering of sites. We sort Kagome sites first by their $y$-coordinate and then by their $x$-coordinate within each row, applying a snake ordering in which the $x$-direction is reversed on every other row \cite{DMRG_snake}. This choice reduces the typical span of interactions in the 1D chain, which helps limit the growth of entanglement that must be represented. The XXZ Hamiltonian is then expressed as a Matrix Product Operator (MPO) directly from the list of nearest-neighbor bonds, constructed using the \texttt{SparseOperatorBuilder} from the \texttt{quimb} library~\cite{gray2018quimb} with a greedy compression scheme. The ground state is found using the two-site DMRG variant (\texttt{DMRG2}), initialized from a random MPS with bond dimension $\chi_0 = 8$ and optimized over the same bond dimension schedule used for the 1D case, running up to 10 sweeps per stage with a convergence threshold of $10^{-6}\,J$.

Despite these mitigations, the 2D mapping inevitably introduces $O(L)$ long-range bonds across the chain, where $L$ is the linear system size. For a geometrically frustrated lattice like Kagome, the ground state carries substantial long-range entanglement, and the required bond dimension grows exponentially with system width. For the $\geq48$-site Kagome lattices studied here, the DMRG calculation did not reach the convergence threshold within the allotted sweep budget at $\chi_{\max} = 160$. Because this calculation is not fully converged, this value is best understood as an upper bound on the true ground-state energy rather than a certified benchmark, and we treat it accordingly when comparing against our SKQD results.

\section{Results}

We apply SKQD to three lattice geometries (a 1D chain, a $2\times N\times3$ Kagome lattice, and a $3\times N$ $J_1$--$J_2$ square lattice) at system sizes ranging from 12 to 72 spins. The chain results serve primarily as a validation against DMRG, for which essentially exact results are available in 1D. The Kagome and square geometries are motivated by their relevance to the QSL phase \cite{square_QSL_evidence, kagome_QSL}. For systems of up to 24 spins, we benchmark against exact diagonalization; for larger systems we use DMRG as described in Sec.~\ref{sec:methods}.

Throughout this section, all results use the anisotropic XXZ model with $\Delta = 2$. As discussed in the supplementary material, the enhanced $ZZ$ coupling produces a sparser ground state in the computational basis, which improves the effectiveness of both the Krylov subspace sampling and the configuration recovery step. For the square lattice we set $J_2/J_1 = 0.5$, placing the model near the expected quantum triple point for this parameter regime \cite{square_QSL_evidence}. No $J_2$ interaction is included for the Kagome or chain geometries.

\begin{figure*}[t]
    \centering
    \includegraphics[width=\textwidth]{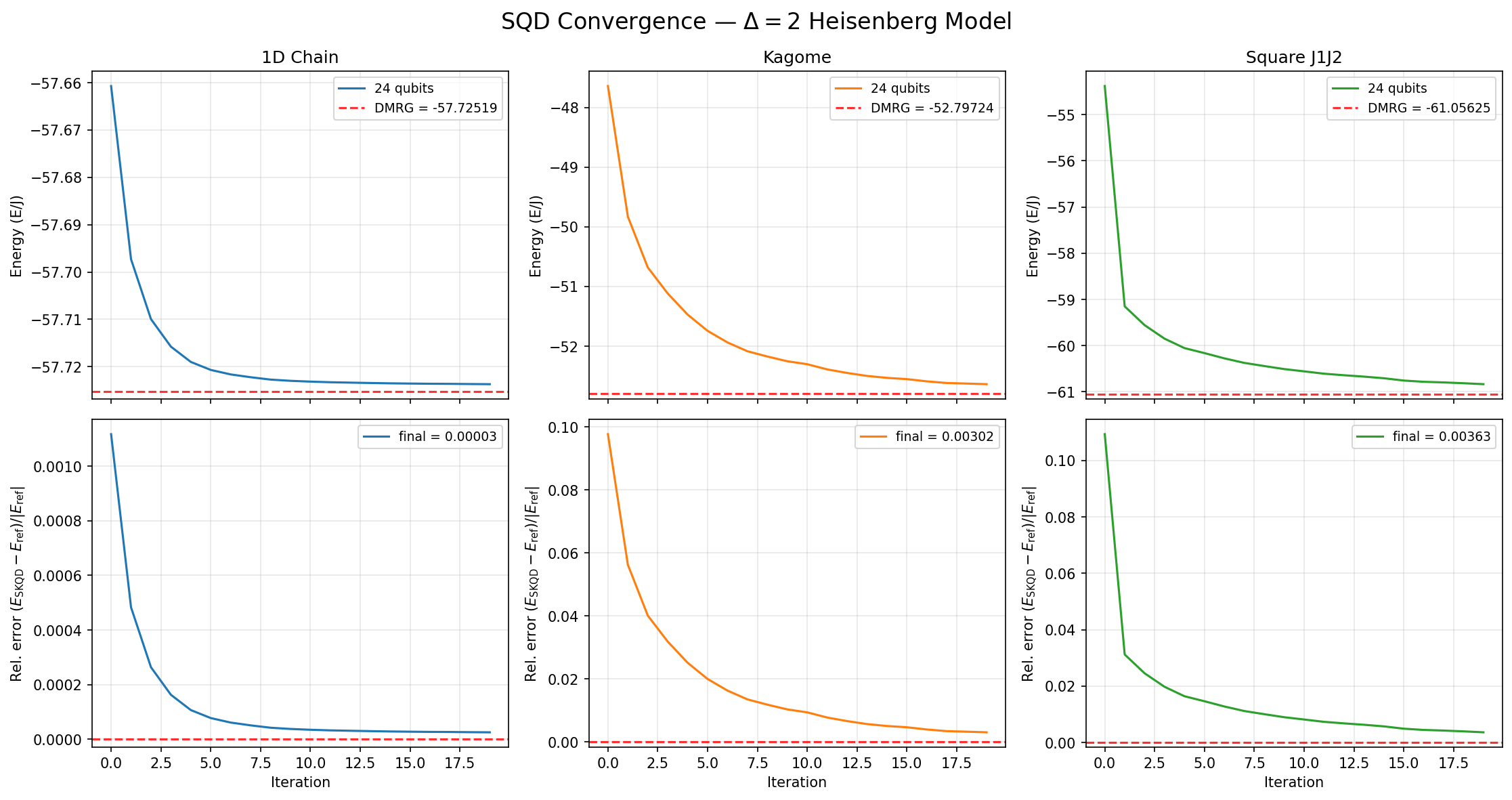}
\caption{SKQD convergence for the $\Delta = 2$ XXZ Heisenberg model on 24-qubit systems across three lattice geometries: the 1D chain ($k{=}20$ Trotter steps), Kagome $2\times4\times3$ lattice ($k{=}5$), and $J_1$--$J_2$ $3\times8$ square lattice ($k{=}5$). \textbf{Top row:} Energy estimates (in units of $E/J$) as a function of SKQD iteration, with dashed lines indicating the exact-diagonalization reference. \textbf{Bottom row:} Corresponding relative error $|E_{\mathrm{SKQD}} - E_{\mathrm{ref}}|/|E_{\mathrm{ref}}|$. SKQD converges monotonically to sub-percent accuracy across all three geometries. All circuits were executed on \texttt{ibm\_pittsburgh} (Heron~r3) with $10^6$ total shots per run.}
    \label{fig:big_energy_plot}
\end{figure*}

Figure~\ref{fig:big_energy_plot} shows the SKQD energy estimate and relative error as a function of iteration number for the 24-qubit instance of each geometry, where exact-diagonalization references are available. In all three cases, the algorithm converges monotonically within the first several iterations, reaching converged relative errors of $3\times 10^{-5}$ (chain), $3.0\times 10^{-3}$ (Kagome), and $3.6\times 10^{-3}$ (square lattice), confirming reliable sub-percent recovery of the ground state at this system size across all three geometries. As shown in Table~\ref{tab:final_err}, performance degrades at larger system sizes across all geometries, with errors reaching $19\%$ for the 72-spin chain, $36\%$ for the 72-site Kagome, and $29\%$ for the 72-site square lattice. Despite being limited to only $k=5$ Trotter steps by its higher gate complexity, the square lattice follows the same general pattern as the other geometries: sub-percent accuracy at small system sizes, with degradation setting in as the ground state density grows and the shallow Krylov subspace becomes insufficient to adequately sample the relevant configurations.

Table~\ref{tab:final_err} summarizes the converged energy estimates and relative errors for all system sizes and geometries. The chain and Kagome results at 12 and 24 sites confirm that SKQD can recover ground-state energies to within $0.4\%$ of the exact reference when the Krylov subspace is sufficiently expressive. Errors at 36 sites and beyond reflect both the growth in ground-state density as the thermodynamic limit is approached, and the increasing mismatch between the available circuit depth and the complexity required to adequately sample the relevant subspace. 

\begin{table*}[t]
\centering
\begin{ruledtabular}
\begin{tabular}{l c c c c c}
\toprule
\textbf{Lattice} & \textbf{Geometry} & \textbf{Qubits} & \textbf{$E_{\mathrm{SKQD}}$ (E/J)} & \textbf{$E_{\mathrm{ref}}$ (E/J)} & \textbf{Error (\%)} \\
\midrule
  Chain  & 12 & 12 & $-28.22782$ & $-28.22782$ & 0.0000 \\
  Chain  & 24 & 24 & $-57.72370$ & $-57.72519$ & 0.0026 \\
  Chain  & 36 & 36 & $-85.42173$ & $-87.30643$ & 2.16 \\
  Chain  & 48 & 48 & $-101.48657$ & $-116.91656$ & 13.20 \\
  Chain  & 60 & 60 & $-122.01879$ & $-146.53704$ & 16.73 \\
  Chain  & 72 & 72 & $-142.05853$ & $-176.16268$ & 19.36 \\
  \midrule
  Kagome  & $2 \times 2 \times 3$ & 12 & $-25.26432$ & $-25.26482$ & 0.0020 \\
  Kagome  & $2 \times 4 \times 3$ & 24 & $-52.63769$ & $-52.79724$ & 0.3022 \\
  Kagome  & $2 \times 6 \times 3$ & 36 & $-65.69712$ & $-80.26089$ & 18.15 \\
  Kagome  & $2 \times 8 \times 3$ & 48 & $-79.59494$ & $-107.70851$ & 26.10 \\
  Kagome  & $2 \times 10 \times 3$ & 60 & $-94.06507$ & $-135.15758$ & 30.40 \\
  Kagome  & $2 \times 12 \times 3$ & 72 & $-103.77176$ & $-162.60706$ & 36.18 \\
  \midrule
  Square  & $3 \times 4$ & 12 & $-29.86216$ & $-29.97930$ & 0.3907 \\
  Square  & $3 \times 8$ & 24 & $-60.83437$ & $-61.05625$ & 0.3634 \\
  Square  & $3 \times 12$ & 36 & $-77.74304$ & $-91.42101$ & 14.96 \\
  Square  & $3 \times 16$ & 48 & $-95.32947$ & $-121.66040$ & 21.64 \\
  Square  & $3 \times 20$ & 60 & $-117.40767$ & $-151.54182$ & 22.52 \\
  Square  & $3 \times 24$ & 72 & $-129.47049$ & $-181.49140$ & 28.66 \\
\bottomrule
\end{tabular}
\end{ruledtabular}
\caption{Ground-state energy estimates from SKQD compared to reference energies (exact diagonalization for systems of $\leq 24$ spins; DMRG otherwise) for the $\Delta = 2$ XXZ Heisenberg model across chain, Kagome, and square lattice geometries. Energy error is computed as $|E_{\mathrm{SKQD}} - E_{\mathrm{ref}}|/|E_{\mathrm{ref}}| \times 100\%$. All quantum circuits were executed on \texttt{ibm\_pittsburgh} (Heron~r3) with $10^6$ total shots. The 1D chain uses $k{=}20$ Trotter steps; Kagome and square lattices both use $k{=}5$.}
\label{tab:final_err}
\end{table*}

\subsection{Comparison with Previous VQE Approaches}

The results above place SKQD in the context of prior quantum algorithm approaches to the frustrated Heisenberg model. The most directly comparable prior work is Weaving et al., who applied the contextual subspace VQE to the antiferromagnetic Kagome lattice and achieved a relative error of $0.01\%$ on a 12-site system \cite{kagome_12spin}. Our SKQD calculation on the same geometry achieves $0.002\%$, surpassing this accuracy while requiring no variational optimization. More importantly, the VQE approach does not extend beyond this system size: the required circuit depth grows rapidly with qubit count, and the resulting accumulation of hardware noise causes energy estimates to increase significantly, a well-documented limitation of the method \cite{VQE_review}. Shiels (2024) applied VQE to the $J_1$--$J_2$ square Heisenberg lattice and limited his study to 16 spins due to energy estimates being sufficiently unreliable. 
To compensate for this, the study shifted focus to alternative observables \cite{Shiels_16spin_square}. 
In contrast, SKQD operates at 24 and 36 sites on the same geometry, albeit with larger errors, which reflect the constrained Krylov depth available for this Hamiltonian. The key advantage of SKQD over VQE is therefore scalability rather than per-site accuracy: by decoupling the quantum sampling step from the classical diagonalization, SKQD sidesteps the barren plateau and noise accumulation problems that constrain VQE circuit depth \cite{VQE_review}, allowing it to be applied to meaningfully larger systems on current hardware.

\subsection{Shot Count Dependence and Convergence Scaling}

\begin{figure}[h]
    \centering
    \includegraphics[width=1\linewidth]{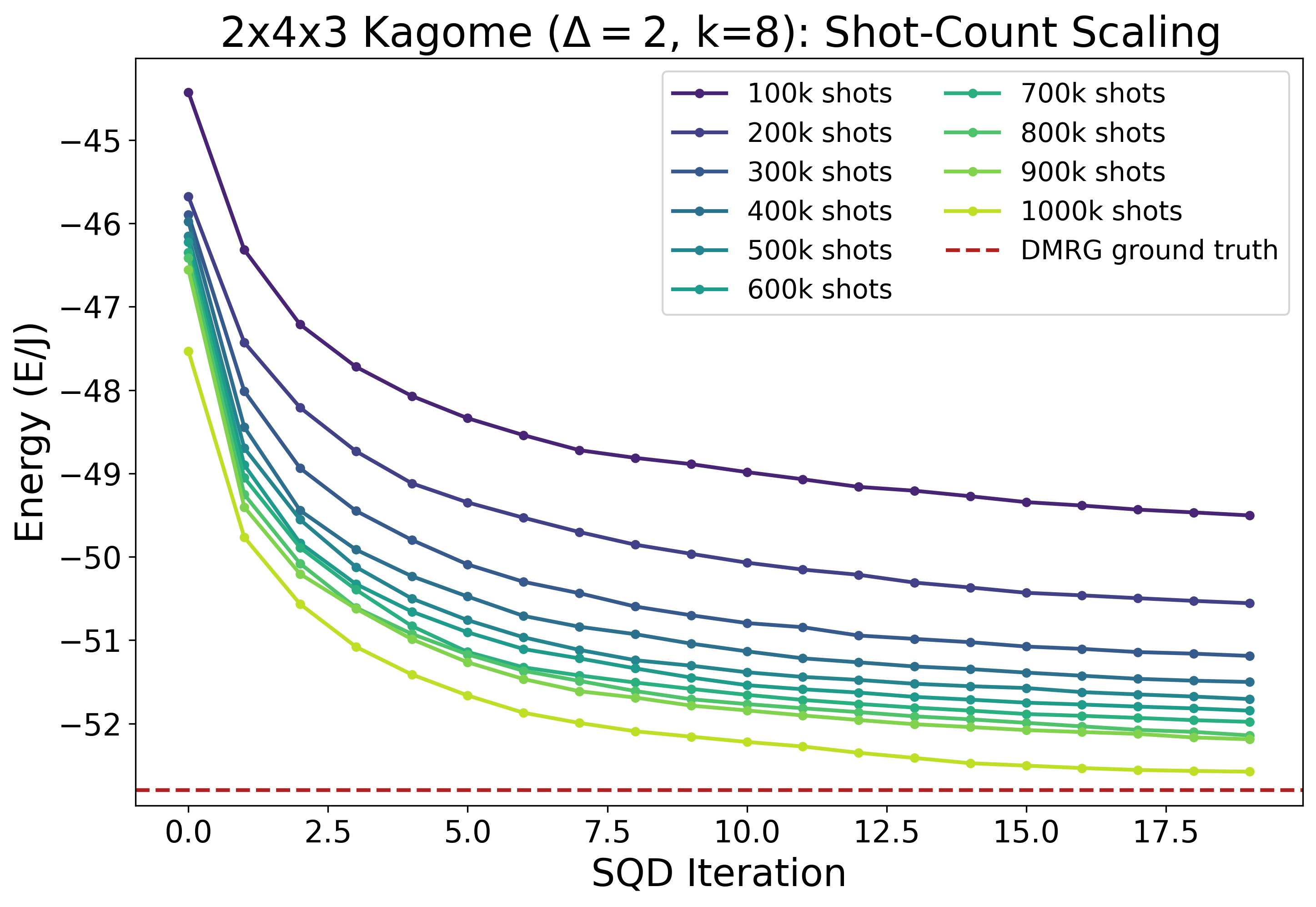}
    \caption{SKQD energy estimates for the $2\times4\times3$ Kagome lattice (24 sites, $\Delta = 2$, $k = 8$ Krylov steps) as a function of iteration number for total shot counts ranging from 100k to 1000k. The red dashed line indicates the DMRG ground-state reference energy. Convergence improves monotonically with shot count, and the 1000k curve has not fully plateaued at 20 iterations, indicating that further increases in  shot count would yield additional accuracy gains.}
    \label{fig:shot_scaling}
\end{figure}

Figure~\ref{fig:shot_scaling} shows SKQD energy estimates for the 24-site Kagome lattice as a function of iteration number across a range of total shot counts from 100k to 1000k, with $k = 8$ Krylov steps on \texttt{ibm\_pittsburgh}. The results exhibit a clear and monotonic improvement with increasing shot count: each additional increment in shots produces a lower converged energy, with the 1000k-shot curve approaching within approximately $1\%$ of the DMRG reference. Importantly, the convergence curves have not fully plateaued at 20 iterations even at 1000k shots, indicating that the algorithm has not yet exhausted the information available in the sampled subspace. This suggests that further increases in shot count or iterations would yield continued accuracy improvements, and that the results reported in Table~\ref{tab:final_err} at $10^6$ total shots represent a conservative estimate of what SKQD can achieve on this geometry.

This shot-count dependence has direct implications for scaling to larger system sizes. At 24 sites, the ground state remains sparse enough so that $10^6$ shots provide near-adequate subspace coverage; at 36 sites and beyond, the denser ground state requires proportionally more measurements to achieve the same coverage, and the failure to converge seen in Table~\ref{tab:final_err} at those sizes is consistent with an insufficient shot budget rather than a fundamental limitation of the algorithm. Stronger classical resources which are capable of projecting and diagonalizing the Hamiltonian in large subspaces created from increases shot count will extend the regime of accurate SKQD estimation to larger system sizes.

\subsection{Initial-State Comparison}

The choice of initial state for Krylov circuit construction directly 
shapes which configurations dominate the sampled subspace. The dimer 
(singlet) state is a natural baseline since it lies in the $S_z = 0$ 
sector and is the exact ground state of the two-site Heisenberg 
Hamiltonian, making it well-aligned with the antiferromagnetic XXZ 
ground state at small length scales. However, a single initial state 
limits subspace diversity, particularly at larger system sizes where 
the ground state spreads across many bitstrings. To address this, we 
augment the dimer prep with a second, complementary initial state: a 
mean-field warm start obtained by coordinate-descent optimization of 
single-qubit Bloch angles $(\theta_i, \phi_i)$ over a discrete grid, 
implemented as a depth-2 product-state circuit ($R_z R_y$ per qubit). 
For each Krylov order $k$, two circuits are submitted, one with each 
prep, and the resulting bitstrings are pooled before configuration 
recovery. The total shot budget is held fixed at $10^6$ across both 
preps, so the warm-start addition contributes no quantum overhead and 
no additional 2-qubit gates.

\begin{figure}[h]
    \centering
    \includegraphics[width=0.9\linewidth]{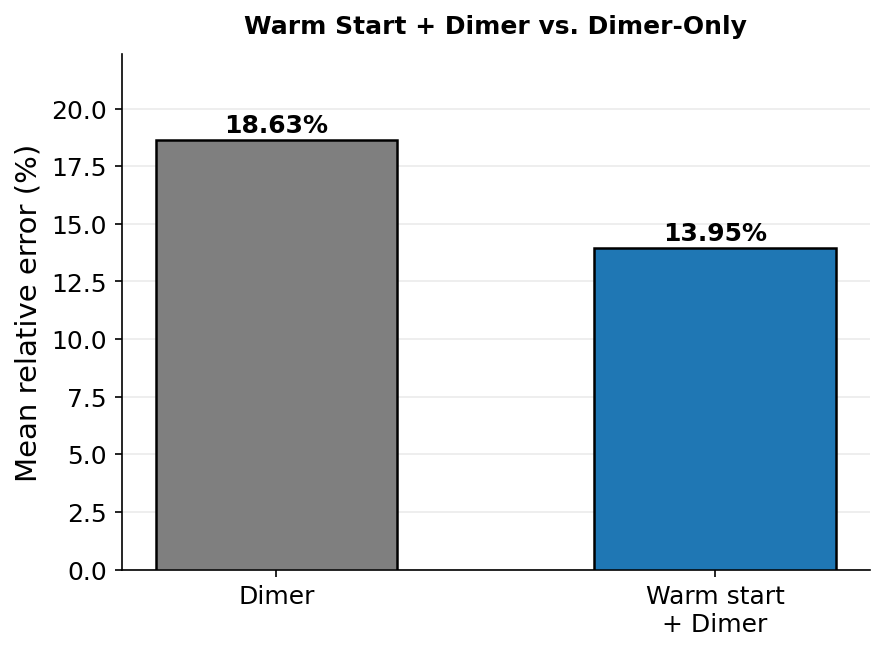}
    \caption{Mean final relative error across all (geometry, system size) points, comparing dimer-only against dimer + warm-start initial-state preparation. Warm start lowers mean final error by 4.68 percentage points (25\% relative reduction)}
    \label{fig:initial_state_comparison}
\end{figure}

Figure~\ref{fig:initial_state_comparison} compares the converged final 
error of dimer-only runs against dimer + warm-start runs across all 
geometries and system sizes on \texttt{ibm\_pittsburgh}. The mean final 
relative error drops from $18.63\%$ (dimer-only) to $13.95\%$ 
(dimer + warm-start), 
a $4.68\%$ absolute and $25\%$ relative 
reduction. The benefit is regime-dependent: at 12--24 qubits, both 
methods recover the reference energy to well under $1\%$ and are 
indistinguishable; at 36 qubits the comparison is mixed, with the chain 
benefiting and the 2D lattices roughly tied; and at 48--72 qubits the 
warm-start addition wins decisively across every geometry, with the 
gap generally growing with system size. The pattern is consistent 
with the picture that subspace diversity matters most precisely where 
the dimer-only run is weakest, namely the larger systems whose denser 
ground states cannot be adequately covered from a single state-prep 
ansatz at fixed shot budget.

\subsection{Two-Qubit Depth and Gate Scaling}

The 2-qubit gate depth is an effective proxy for circuit feasibility on near-term hardware: 2-qubit gates are the dominant source of noise, and depth governs the accumulation of errors over the circuit. On Eagle-class processors, fidelity bounds imply that circuits with $\sim$100 qubits and $\sim$100 gate layers already yield state fidelities below $5 \times 10^{-4}$ \cite{Kim2023utility}; the lower error rates of the Heron family extend this limit substantially. We impose a gate-depth budget of 600 as a conservative bound on reliable execution throughout this work.

\begin{figure}[h]
    \centering
    \includegraphics[width=1\linewidth]{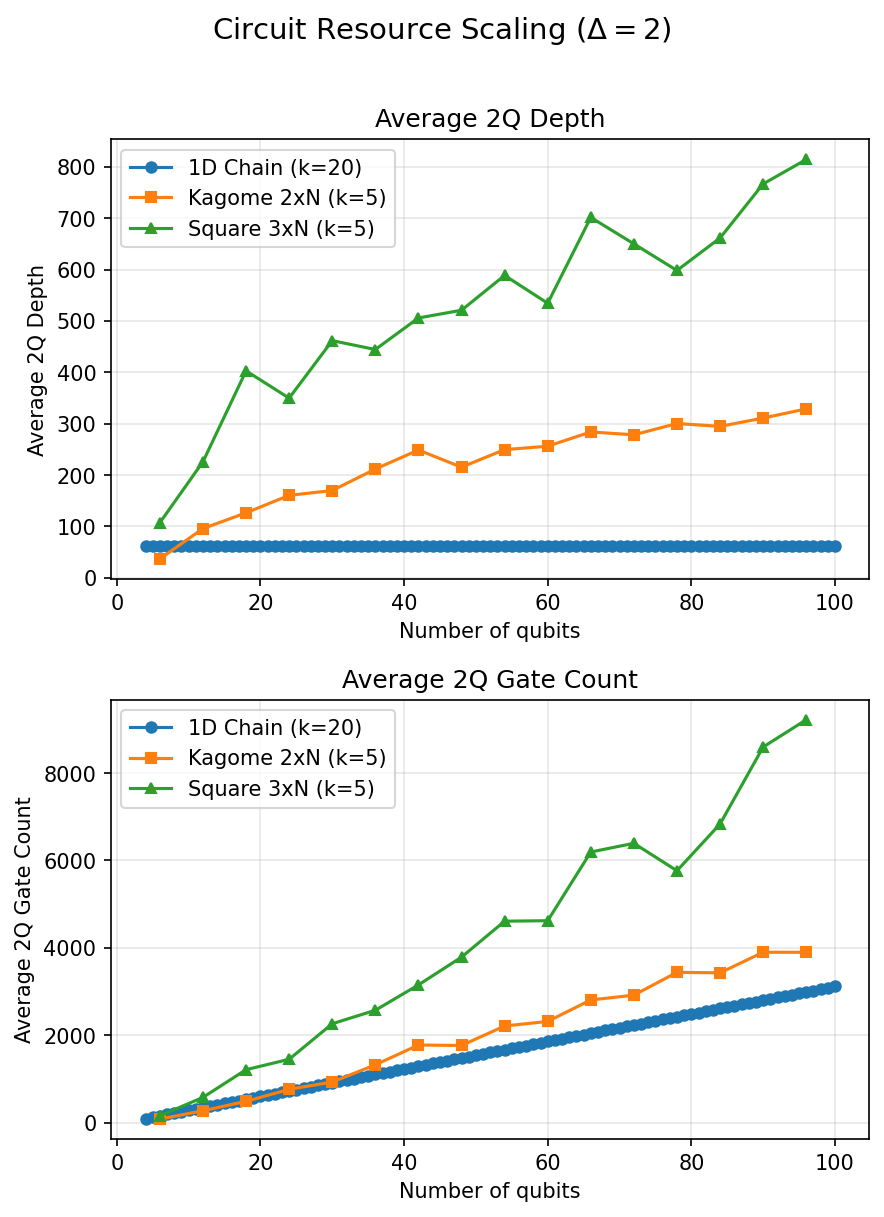}
    \caption{Average 2-qubit gate depth (top) and gate count (bottom) as a function of system size for the 1D chain ($k{=}20$), Kagome $2{\times}N$ ($k{=}5$), and square $3{\times}N$ ($k{=}5$) lattices.}
    \label{fig:depth_scaling}
\end{figure}

\begin{figure}[h]
    \centering
    \includegraphics[width=1\linewidth]{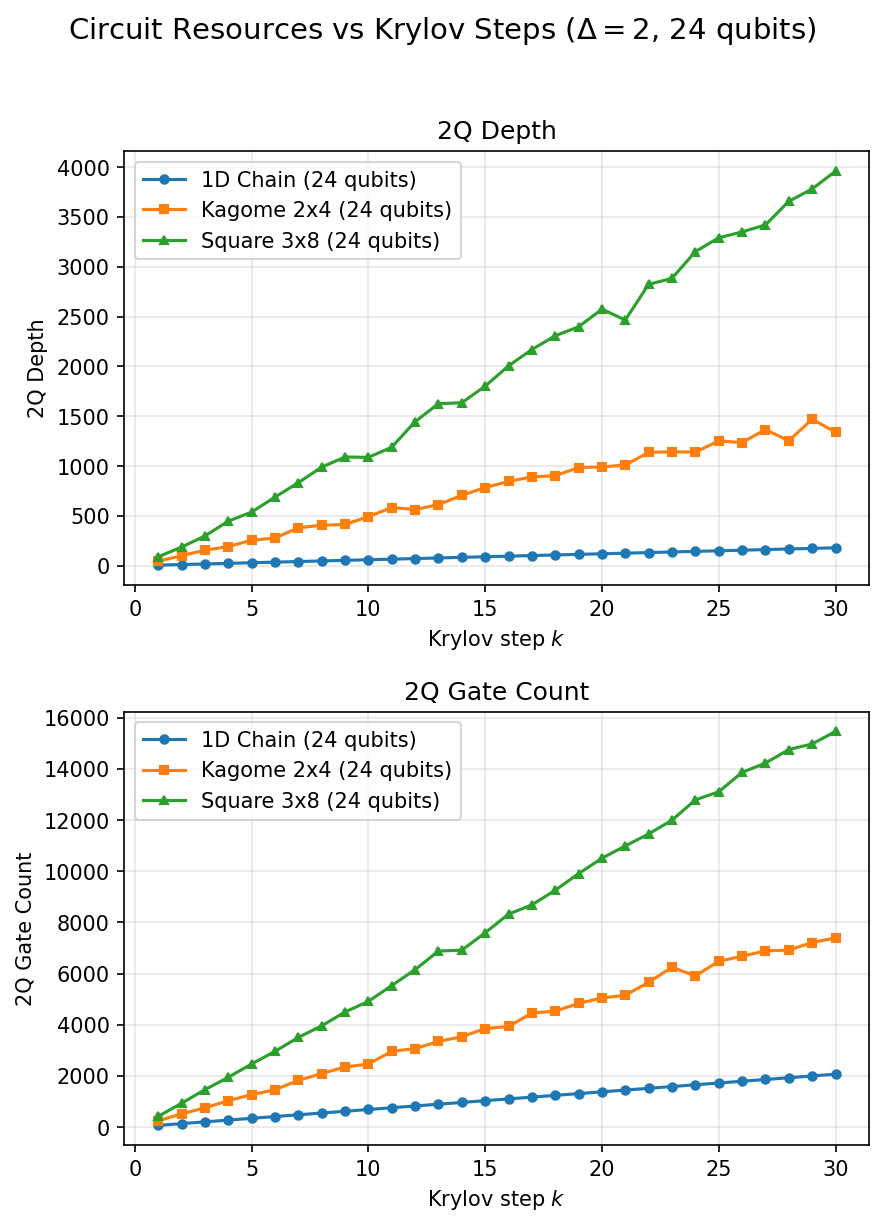} \caption{2-qubit gate depth (top) and gate count (bottom) as a function of Krylov step $k$ for 24-qubit systems, comparing the 1D chain, Kagome $2{\times}4{\times}3$, and square $3{\times}8$ lattices.}
    \label{fig:krylov_scaling}
\end{figure}

To characterize resource scaling, we compiled Krylov circuits for each geometry across a range of qubit counts and values of $k$ ($\Delta = 2$), transpiled to the 156-qubit Heavy Hex topology of \texttt{ibm\_pittsburgh} using 10 seeds and selecting the lowest-depth result. Figure~\ref{fig:depth_scaling} shows scaling with system size at fixed Krylov order; Figure~\ref{fig:krylov_scaling} shows scaling with $k$ at 24 qubits.

The results reveal a clear hierarchy. The 1D chain is the most efficient: its average depth remains essentially constant with system size due to the chain Hamiltonian's natural compatibility with Heavy Hex, while gate count grows only linearly. The Kagome lattice scales moderately, staying within the depth budget across all sizes studied. The square lattice is the most demanding, with both depth and gate count growing substantially --- a consequence of the dense connectivity introduced by competing $J_1$ and $J_2$ interactions.

Both metrics grow approximately linearly with $k$ for all geometries, as each Krylov step appends one Trotter block. The per-step overhead is far larger for the square lattice, limiting it to $k = 5$ within our approximate depth budget of 600. This overhead stems from the $J_2$ interactions requiring additional SWAP operations on the Heavy Hex topology, an architectural mismatch that would be substantially reduced on a square-lattice processor such as IBM's Nighthawk, potentially enabling deeper Krylov expansions and improved SKQD accuracy on this geometry.

\subsection{Analysis}

For the 1D chain, Kagome lattice, and square lattice, SKQD achieves sub-percent relative error at 12 and 24 sites, confirming reliable convergence across all three geometries in regimes where exact benchmarks are available. Performance begins to degrade at larger system sizes, though the onset varies by geometry. The 1D chain is the most resilient: errors remain at $2.2\%$ at 36 sites before climbing to $13\%$, $17\%$, and $19\%$ at 48, 60, and 72 sites respectively. The Kagome lattice degrades more sharply, jumping from sub-percent error at 24 sites to $18\%$ at 36 sites and reaching $36\%$ at 72 sites. In both cases, this drop-off reflects the growth of ground-state density as the system grows, which increases the number of bitstrings required to adequately represent the ground state and places greater demands on both the circuit sampling and configuration recovery steps.

The square lattice follows the same general pattern, though degradation sets in earlier arising from increased circuit depth caused by the complexity introduced by the $J_2$ interactions. Sub-percent accuracy is achieved at 12 and 24 sites despite the Hamiltonian being restricted to only $k = 5$ Krylov steps by our gate-depth budget of 600, indicating that even a shallow subspace is sufficient to capture the low-energy structure at these system sizes. Error climbs to $15\%$ at 36 sites, where the combination of growing ground-state density and limited subspace expressibility at $k = 5$ prevents adequate convergence. The 36-site result should therefore be interpreted as a quantum and classical hardware-limited lower bound on SKQD performance for this geometry rather than a ceiling of the algorithm. The classical limitation arises from the eigensolver step of SQD, where the addition of more bitstrings/shots leads to a more memory-intensive calculation. This can be remedied with stronger classical resources. The quantum limitation arises from the noise associated with additional Krylov steps, and improvements in this case are expected as lower gate error rates or square-lattice-native processors become available.

\section{Conclusion}

We have demonstrated the application of SKQD to the antiferromagnetic XXZ Heisenberg model across three geometries (a 1D chain, the Kagome lattice, and the $J_1$--$J_2$ square lattice), spanning system sizes from 12 to 72 spins. Two modifications to the standard SKQD framework were introduced to improve performance for spin models. First, a canonical bitstring compression scheme resolves the conflict between spin-flip degeneracy and configuration recovery, preventing the average occupancy vector from collapsing to a trivial $[0.5, 0.5, \dots]$ array and restoring the algorithm's ability to extract information from noisy circuit outputs. Second, combining measurement data from multiple distinct initial states (a dimer prep and a mean-field warm start) improves Krylov subspace coverage and accelerates convergence without any increase in quantum resources. The combination of Krylov subspace sampling, spin-sector filtering, and degeneracy expansion proves effective in the $\Delta = 2$ XXZ regime, where the enhanced $ZZ$ coupling yields a sparse ground state that is well-suited to bitstring-based diagonalization.

SKQD achieves sub-percent relative error at system sizes up to 24 spins across all three geometries, including a ground-state energy error of $0.002\%$ on the 12-site Kagome lattice. This surpasses the best prior quantum result on this geometry, a VQE error of $0.01\%$, while operating without variational optimization and scaling to larger systems than VQE can access. Notably, the square lattice achieves sub-percent accuracy at 12 and 24 sites despite being restricted to only $k = 5$ Krylov steps by its higher gate complexity, demonstrating that SKQD performs well even under significant hardware constraints at moderate system sizes. At larger system sizes, accuracy degrades across all geometries due to increasing ground-state density and the growing mismatch between available circuit depth and the complexity required to adequately sample the relevant subspace.

While classical tensor network methods such as DMRG remain state-of-the-art for the frustrated Heisenberg model, this work establishes SKQD as the most accurate and scalable quantum algorithm demonstrated for this problem to date. The accuracy and system sizes achieved here exceed those of prior VQE studies by a significant margin, and the methods introduced are broadly applicable to any spin model with a global discrete symmetry.\\

\section{Outlook}
The scalability of SKQD is expected to improve substantially as quantum hardware matures. Reductions in 2-qubit gate error rates will allow deeper Krylov circuits, directly benefiting geometries with high-connectivity Hamiltonians such as the $J_1$--$J_2$ square lattice. Increased shot counts will improve subspace coverage and reduce statistical noise in the configuration recovery step, extending the system sizes at which sub-percent accuracy is achievable for both the chain and Kagome geometries.

The most significant near-term opportunity is the inclusion of periodic boundary conditions. These were excluded in this work due to the long-range qubit connectivity they require and the associated noise penalty on current devices, but they are of particular interest for QSL studies: open boundary conditions introduce edge effects that can obscure bulk phase behavior, and finite-size extrapolation to the thermodynamic limit is more reliable from periodic systems. Beyond the geometries studied here, SKQD is a natural candidate for application to other frustrated lattices relevant to QSL physics, including the triangular and pyrochlore lattices, as well as to models with longer-range interactions or anisotropic couplings.

\section{Acknowledgments}
This work is supported by the NSF CAREER award under grant number 2044842 and the RPI School of Science Computational Excellence Fund. The authors would like to acknowledge Nate Earnest-Noble from IBM Quantum for helpful discussions and technical insights.

\bibliographystyle{apsrev4-2}
\bibliography{refs}

\clearpage
\onecolumngrid
\appendix
\section*{Supplemental Material}
\FloatBarrier

\section{Discussion of Sparsity}

\begin{figure}
    \centering
    \includegraphics[width=0.8\linewidth]{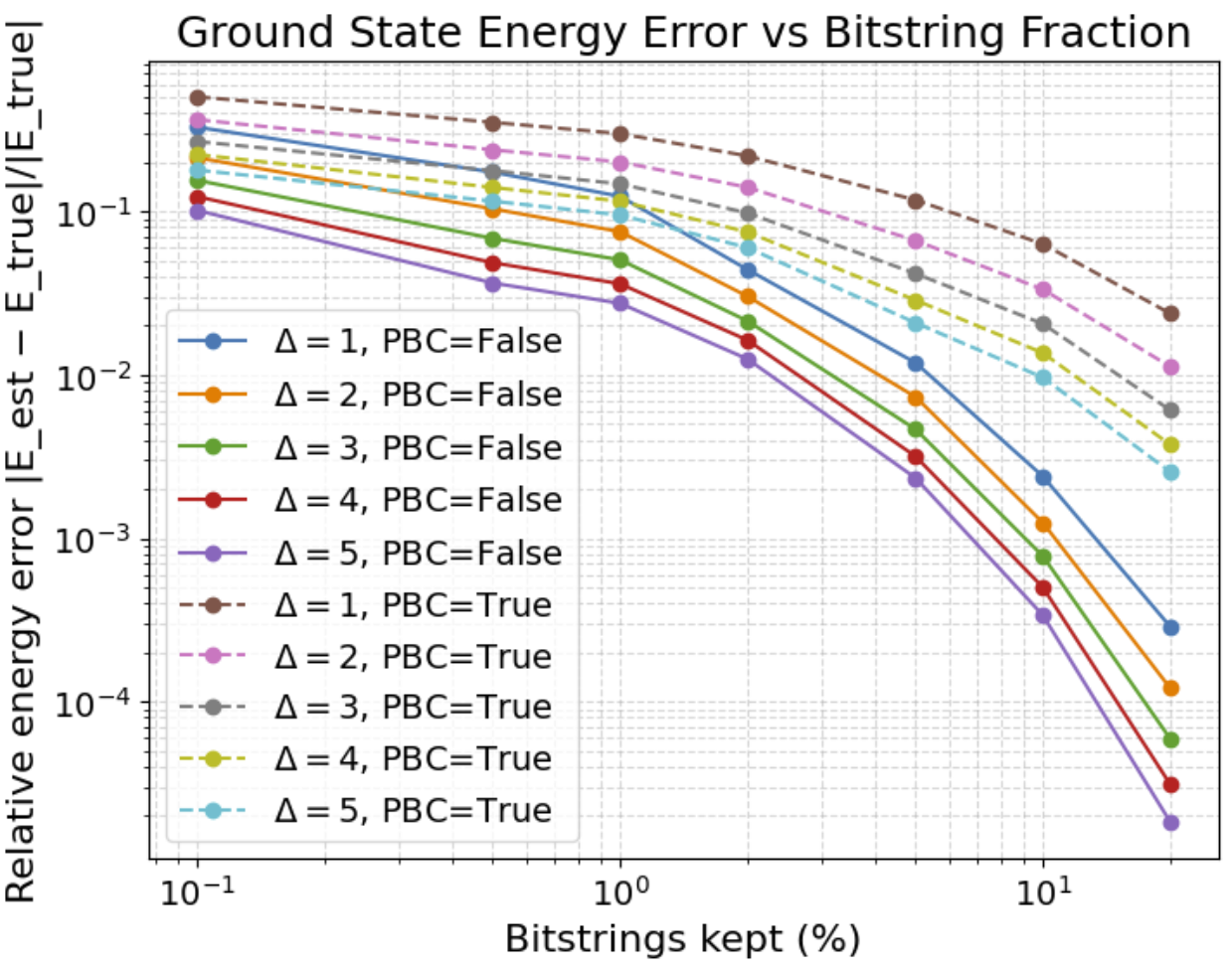}
    \caption{Illustration of ground state sparsity for different values of the anisotropy term $\Delta$ and different boundary conditions. The horizontal axis shows \% of total subspace used for ground state estimate and the vertical axis shows error relative to ground truth.}
    \label{fig:sparsity}
\end{figure}

\begin{figure}
    \centering
    \includegraphics[width=0.8\linewidth]{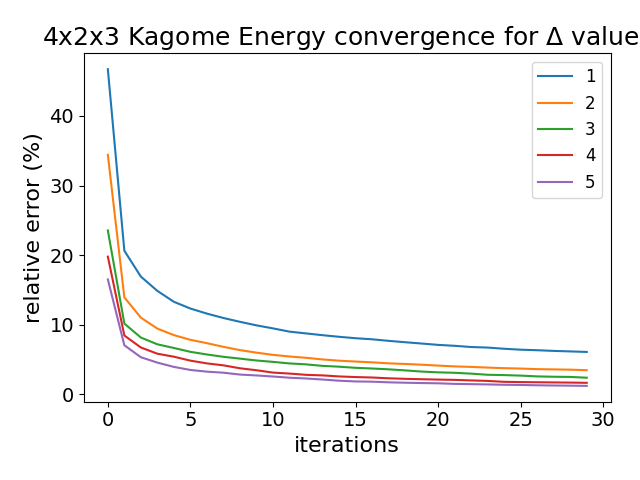}
    \caption{Energy convergence of SKQD for the 4x2x3 Kagome lattice with different values for the anisotropy factor $\Delta$.}
    \label{fig:delta_convergence}
\end{figure}

While the total Hilbert size scales with $2^N$, with N being the number of lattice sites, we know the ground state of the antiferromagnetic model will have total spin 0. This means the ground state will lie entirely in a truncated Hilbert space of dimension $N \text{ choose } N/2$. The sparsity of the ground state in this reduced Hilbert space is an important factor in the efficiency of SKQD. Figure \ref{fig:sparsity} shows how a smaller value of $\Delta$ results in a less sparse ground state, illustrated by the need for more of the subspace to get an accurate ground state estimate. This plot was generated by exactly solving the 2x3x3 (18 spin) Kagome lattice in the Z basis to obtain a ground state eigenvector. The elements of this vector act as weights for every bitstring in the reduced Hilbert space. We then take the top k\% of bitstrings by weight and use these to project the Hamiltonian and estimate the ground state, mirroring the process described for SKQD above. Figure \ref{fig:delta_convergence} shows this in practice after applying SKQD to the 2x4x3 (24 spin) Kagome lattice on quantum hardware. Smaller values for $\Delta$ are less sparse and therefore more difficult to estimate. The relative energy error is close to 1\% for $\Delta=5$ but increases to almost 10\% for $\Delta=1$. 

\section{Effect of SKQD Modifications}

\begin{figure}
    \centering
    \includegraphics[width=0.8\linewidth]{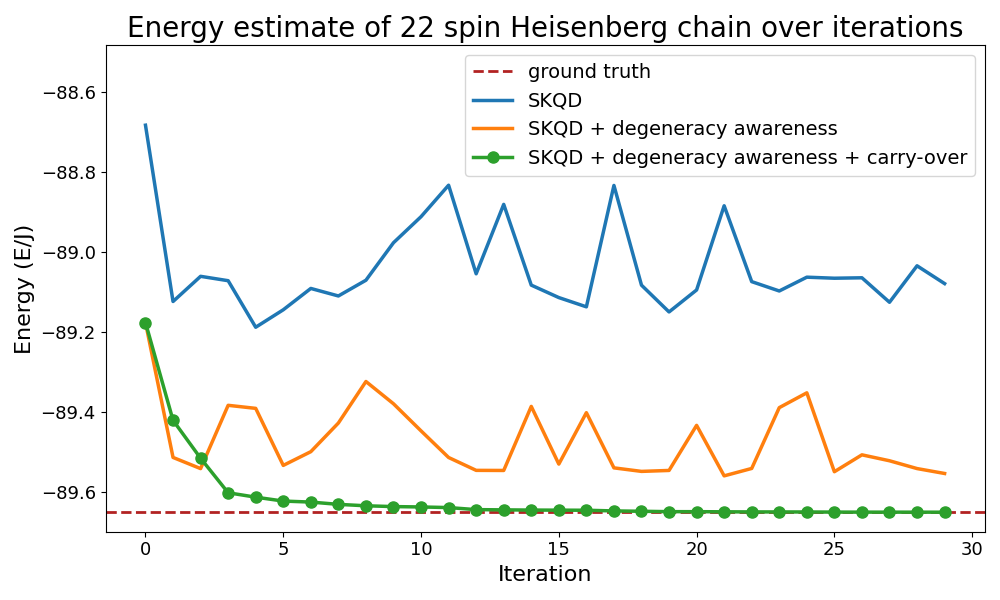}
    \caption{Energy convergence of SKQD for a 22 spin chain. The blue line indicates the original SKQD algorithm, the orange line indicates SKQD with the addition of the degeneracy awareness discussed in the main paper. This includes the compression and expansion of the dataset before and after configuration recovery. The green line indicates convergence for SKQD with degeneracy awareness and the carry-over bitstring system described in the main paper. }
    \label{fig:skqd_effects}
\end{figure}

In the main paper we discuss several modifications which are made to the SKQD algorithm to improve results. Figure \ref{fig:skqd_effects} illustrates the effect of these modifications. With the original SKQD algorithm, the convergence is not strong enough to reach the ground truth. The degeneracy awareness system consists of expanding the bitstring dataset to include all bitwise complements. This increases the subspace coverage and improves the ground state estimate, seen as the orange line being closer to ground truth than the blue line. To facilitate convergence, the "carry-over" system is then added, where the top 30\% of recovered bitstrings from the previous configuration recovery iteration are manually added into the next iteration. This prevents important recovered bitstrings from being lost between iterations due to the probabalistic nature of configuration recovery. The combination of these modifications results in strong convergence to the ground truth. 

\section{Subspace Scaling}

The dimension of the full Hilbert space for an $N$-site spin-$1/2$ 
Heisenberg Hamiltonian scales as $2^N$. For even $N$, the antiferromagnetic 
ground state lies in the total magnetization sector $S_z=0$. In the 
computational basis, this sector consists of bitstrings with exactly $N/2$ 
spin-up sites and $N/2$ spin-down sites. Its dimension is therefore the 
central binomial coefficient,
\[
\dim \mathcal{H}_{S_z=0} = \binom{N}{N/2}.
\]

\begin{figure}
    \centering
    \includegraphics[width=0.8\linewidth]{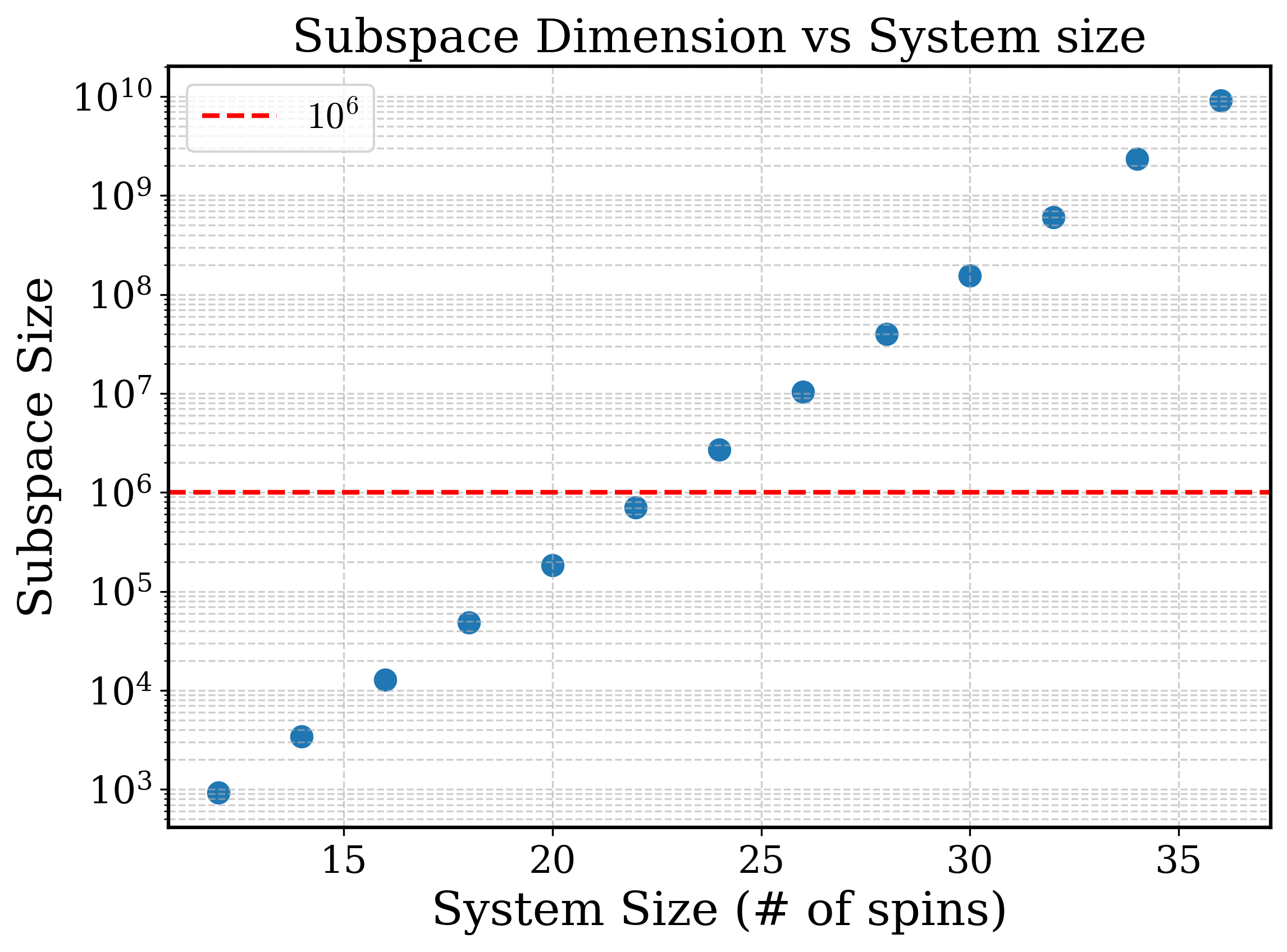}
    \caption{Subspace scaling for the antiferromagnetic Heisenberg ground state. The dimension is equal to $N$ choose $N/2$. A horizontal line has been drawn at $10^6$, indicating the number of quantum computer shots used in our experiments.}
    \label{fig:subspace_scaling}
\end{figure}

Although the $S_z=0$ restriction provides a substantial reduction relative to the full $2^N$ Hilbert space, it does not remove the exponential scaling. The central binomial coefficient grows asymptotically as
\[
\binom{N}{N/2} \sim \frac{2^N}{\sqrt{\pi N/2}},
\]
so this truncated sector still becomes extremely large at modest system sizes. This has direct implications for SKQD. With a fixed shot budget, the number of distinct configurations that can be 
observed is bounded by the number of measurements. Once $\binom{N}{N/2}$ exceeds the shot budget, the sampled configuration set necessarily covers only a small fraction of the symmetry sector. Figure \ref{fig:subspace_scaling} shows the subspace size as a function of system size, with a dashed line drawn at $10^6$, indicating the number of shots used in this work. Systems up to 25 spins are well-covered by this shot count, while larger systems are not. This is observed in the relative energy error table of the main text, where the two-dimensional 12 and 24 spin systems have sub-percent error while systems with $\geq30$ spins have error on the order of $10\%$. More shots would result in more coverage of the subspace, leading to lower error on these larger system sizes. 

\section{Limitations}
The primary limitation of SKQD as applied here is its dependence on ground-state sparsity in the computational basis. The choice of $\Delta = 2$ was motivated by the sparser ground states it produces, making the problem more tractable for bitstring-based methods. As $\Delta$ decreases toward the isotropic XXX point ($\Delta = 1$), the ground state becomes denser, requiring more measurements to achieve the same subspace coverage and making configuration recovery increasingly unreliable. Extending SKQD to the fully isotropic regime will likely require a larger shot budget, a more targeted initial state preparation, or a change of computational basis in which the ground state is naturally sparser.

A secondary limitation is the sensitivity of results to circuit depth and gate fidelity, as illustrated by the square lattice. When the Trotter circuit for a given Hamiltonian saturates the available gate-depth budget at low Krylov order, the sampled bitstrings carry substantial hardware noise and the Krylov subspace is poorly conditioned. In this regime, additional shots help only marginally; what is needed is either lower gate error rates or a more hardware-efficient Trotter decomposition. Finally, for systems beyond 24 spins, the absence of exact diagonalization benchmarks means that our error estimates rely on DMRG reference energies that are themselves approximate, particularly for the $\geq48$-site Kagome lattices where the DMRG calculation did not fully converge.

%\bibliography{references}

\end{document}